# Second harmonic generation from Chalcogenide metasurfaces via mode coupling engineering

Tapajyoti Das Gupta,[a] Louis Martin-Monier,[a] Jeremy Butet,[b] Kuang-Yu Yang,[b] Andreas Leber,[a] Chaoqun Dong,[a] Tung Nguyen-Dang,[a] Wei Yan,[a] Olivier J.F. Martin,[b] Fabien Sorin[a]†

[a] Laboratory of Photonic Materials and Fiber Devices, Ecole Polytechnique Fédérale de Lausanne, Switzerland.

[b] Nanophotonics and Metrology Laboratory, Ecole Polytechnique Fédérale de Lausanne, Switzerland.

† Corresponding Author, email: fabien.sorin@epfl.ch

Abstract: Dielectric metasurfaces have shown prominent applications in nonlinear optics due to strong field enhancement and low dissipation losses at the nanoscale. Chalcogenide glasses are one of the promising materials for the observation of nonlinear effects due to their high intrinsic nonlinearities. Here, we demonstrate, experimentally and theoretically, that significant second harmonic generation can be obtained within amorphous chalcogenide based metasurfaces by relying on the coupling between lattice and particle resonances. We further show that the high quality factor resonance at the origin of the second harmonic generation can be tuned over a wide wavelength range using a simple and versatile fabrication approach. The measured second harmonic intensity is orders of magnitude higher than that from a deposited chalcogenide film, and more than three orders of magnitude higher than conventional plasmonic and Silicon-based structures. Fabricated via a simple and scalable technique, these all-dielectric architectures are ideal candidates for the design of flat non-linear optical components on flexible substrates.

## Introduction

High refractive index dielectric metasurfaces[1–4] are sub-wavelength nanostructured arrays that provide a promising alternative to plasmonic structures[5,6], especially thanks to their intrinsic low dissipation losses[2], multipolar resonances[7] and their inherent magnetic response. An important application of such dielectric resonators lies in nonlinear optics[7]. In particular, second harmonic (SH) generation has thus far mainly relied on III-V semiconductors[8,9] due to their non-centrosymmetric crystalline structures. Other common candidates include silicon[10], plasmonic-based metasurfaces[11,12] and hybrid structures[13] which have all shown a strong SH generation enhancement as compared to their thin films counterpart.

SH signals can also be strongly enhanced by using properly engineered resonances. For example, resonators with broken symmetry[8,14] have proven efficient to produce high quality factor (Q.F.) resonances and hence strong field enhancement at both the fundamental and the SH wavelengths. Other approaches lie in the fabrication of high aspect ratio pillar-like structures[15], where the so-called bound state in continuum allows for a high Q.F. using single pillar structures. Similar high efficiency has also been reported in plasmonic architectures, for example with complex Metal-

Dielectric-Metal (MDM) cavities relying on patch antennas[16] or chiral structures[17]. These interesting designs nevertheless require a high aspect ratio or sophisticated shapes, thus involving complex e-beam processes, which limits the applicability of such devices.

Another way to engineer high Q.F. metasurfaces is to use surface lattice resonances. For a bare lattice made from non-resonant particles, near-field enhancement is observed at the Rayleigh-anomaly wavelength. [21,22]. This resonance spectral position is defined by:

$$\lambda_{(i,0)} = a_0 \cdot \left(\frac{n}{|i|} - \frac{\sin\theta}{i}\right) \tag{i}$$

where $a_0$ is the periodicity or the lattice constant of the arrays, $n$ is the surrounding refractive index, $\theta$ is the angle of incidence, and $i$ is an integer denoting the diffraction order. Under normal incidence, this resonance is hence solely dependent on the lattice periodicity and the surrounding refractive index[18]. By additionally considering resonant nanoparticles, a wealth of optical phenomena and spectral features can be exploited by relying on the coupling between lattice and Mie-type modes associated to the particle[23], as illustrated in Fig. 1.

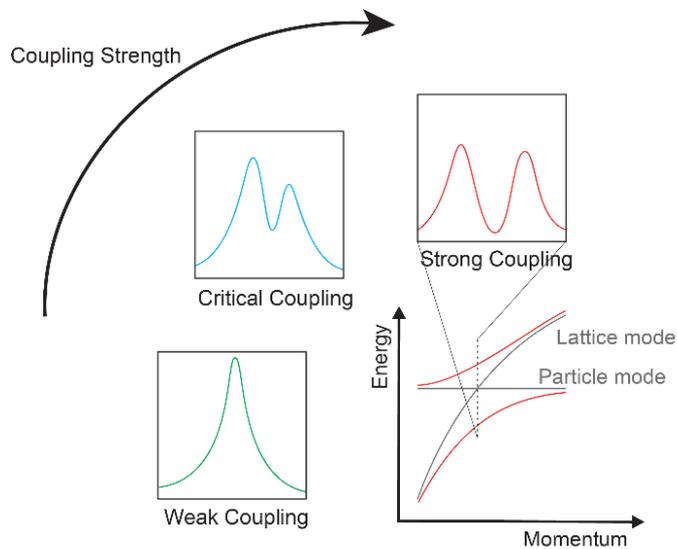

*Figure 1: The optical response of a system that supports both lattice and particles modes can be tuned by adjusting the coupling between both modes. Weak coupling leads to a Lorentzian response (green spectrum), critical coupling produces a Fano-type line shape (blue), while the response of the system exhibits two symmetric peaks in the strong coupling regime (red). These two peaks correspond to the energy splitting associated with the avoided crossing between the lattice and particle modes*

For weak coupling, both resonances are barely modified and the response of the system remains essentially Lorentzian. When coupling reaches the critical regime, this interaction leads to the emergence of a high Q.F. Fano-type resonance[24,25]. A strong coupling regime further leads to an energy splitting yielding two characteristic peaks[26]. Interestingly, the entire response of the system can be controlled by the interactions between particle resonance (determined by its material composition and shape) and the lattice resonance, which is governed by the array periodicity and the surrounding refractive index [27,28]. Simple geometrical levers such as lattice periodicity or interparticle gap

(equivalent to particle size for a fixed period) allow for tunable coupling between particle and lattice modes. The resonator shape or constituting materials further represent additional degrees of freedom. Engineering strong resonance through engineered mode coupling proves a versatile platform for light-matter interaction at the nanoscale, and in particular for enhancing SHG emission[29].

However, independent control over geometrical parameters commonly relies on time and cost-intensive lithographic techniques and thus fails to optimize the nano-structures for enhanced SH performance. Previous works have proposed simple techniques to controllably tune a single geometrical parameter without relying on lithographic techniques, for example using thermoplastics[19], stretched PDMS during nanoimprinting process[30], or using elastomeric PDMS[31] as a substrate. By either using a constant particle size or constant lattice period, these works provided a single geometrical lever to control mode coupling. A simple and versatile way to independently tune two or more geometrical parameters remains to be developed.

Here, we address this issue and demonstrate a novel materials and fabrication platform for high conversion efficiency SH dielectric nanostructures from the coupling between particle resonance with lattice resonance, via the template dewetting of chalcogenide glasses-based metasurfaces. Chalcogenides[32–36] are ideal dielectric for SHG in the proper form and configuration[37,38], yet they have been seldom used due to a lack of proper resonance engineering strategy, and versatile materials processing approach. The process we developed[39] enables tuning independently the lattice period (through imprinting pressure) and the particle size or diameter (through successive dewetting) in order to understand and optimize the emergence of Fano-type surface lattice resonance. With this fabrication flexibility, coupling between the diffractive lattice and the radiative particle modes can be properly controlled, which is subsequently exploited to enhance the SH signal. We observe an enhanced SHG at the Fano-type surface lattice resonance of two orders of magnitude larger than the non-resonant region and four orders of magnitude higher compared to unstructured thin films. The SH conversion efficiency we observe is of the order of $10^{-6}$, which outperforms conventional plasmonics and Si-based structures. This experimental effect is investigated through finite difference time domain (FDTD) simulations by relating the origin of the strong SHG to the field enhancement induced by critical coupling between the lattice and radiative modes. The design methodology and SH efficiency reported in this article pave the way for nonlinear chalcogenide-based light sources using a highly scalable and versatile fabrication platform.

## Experimental Section

*Fabrication Technique*

An array of inverted pyramids with different periods and sizes was obtained by interference lithography followed by KOH anisotropic etching on p-doped Si [100] wafer (CEMITEC, Spain). Using a negative PDMS stamp (Sylgard 184, Dow Corning, USA), the Si microstructure was reproduced via nano-imprinting onto a thermoplastic polycarbonate sheet (Goodfellow, UK). The thin chalcogenide film (ChG) was then deposited by thermal evaporation (UNIVEX 250, Oerlikon, Germany). Annealing was achieved on a hot plate (Isotemp Fisher Scientific) (80°C) to induce thermal dewetting.

*SEM characterization and analysis*

All SEM samples were coated with a 10 nm carbon film. The SEM images were taken with a Zeiss Merlin field emission SEM equipped with a GEMINI II column operating at 3.0 kV with a probe current of 120 pA. All SEM image analysis were done using Image J software, averaging the measures over a large number N of particles for the size distribution (N>100) and a reduced number of periods P for the period extraction (P>5).

*Simulations*

The linear simulations were performed using commercially available FDTD software Lumerical (version 2018 a). The optical constants were obtained experimentally via spectroscopic ellipsometry (Sopra GES 5E) analysis of thin film of selenium (Figure S.I.3). A linearly polarized plane wave source was used for all the simulations. The substrate index was chosen to be 1.6, which is the average index of polycarbonate[40]. Periodic boundary conditions were used in the transverse direction and a perfectly matched layer was used in the wave propagation direction. The particle is defined as the overlap between an ellipsoid with in-plane radius r and conical shape of base length d. The ellipsoid in-plane radius r is defined as $r = d \cdot (1+\sqrt{2})/2$, while the ellipsoid center coincides with the pyramid base center. The particle radius in this work refers to the ellipsoid radius in the model. The height of the pyramid shape was kept fixed at 200 nm. For the electric field amplitude spectral evolution, a time monitor was placed in the interparticle gap i.e. in between two neighbor ellipsoids as well as in the center of the ellipsoidal particle. For magnetic field amplitude spectral evolution, a time monitor was placed at the particle center (i.e the ellipsoid center). For the maximal electric field enhancement calculation, the value of the maximal field intensity was extracted from the electric field distribution map at the corresponding resonance wavelength, within the metasurface's symmetry plane. Additional calculations based on the Mie theory were performed to determine the resonance conditions for Se particles as a function of their radius (Figure S.I.4).

*Linear and Non-linear measurements*

The linear transmission spectra were measured using an integrating sphere coupled with a spectrometer (Ocean Optics, UV-Vis USA). For the non-linear optical measurements, a commercial multiphoton scanning microscope (LEICA SP5MULTI-PHOTON) combined with a 20x/1.00 NA water-immersion objective (HCX PL APO) and a Ti:Sapphire femtosecond laser (Chameleon ultra-laser, repetition rate: 80 MHz, operating wavelength $\lambda = 800$ nm, pulse duration: 140 fs) combined with an electro-optical modulator (EOM) to adjust the input power was used. The backward scattered SHG from the structures was collected by the same objective followed by a beam splitter, a band-pass filter (of different wavelengths with a bandwidth of 10 nm), and a photomultiplier tube (NDD PMT). The XY scanner (400 Hz scanning speed) of the microscope enabled a 387.5 μm × 387.5 μm field range with 758.32 nm pixel resolution at 800 nm. The quadratic dependence between the SH intensity and the input power was recorded be carefully tuning the input power.

## Results and discussion

*Fabrication of metasurfaces*

The fabrication process is schematized in Figure 2(a) and is detailed in the experimental section (Technique of fabrication). The obtained pattern is illustrated in the SEM images in Figure 2(b) (see Figure S.I.1 for additional details on the fabrication method).

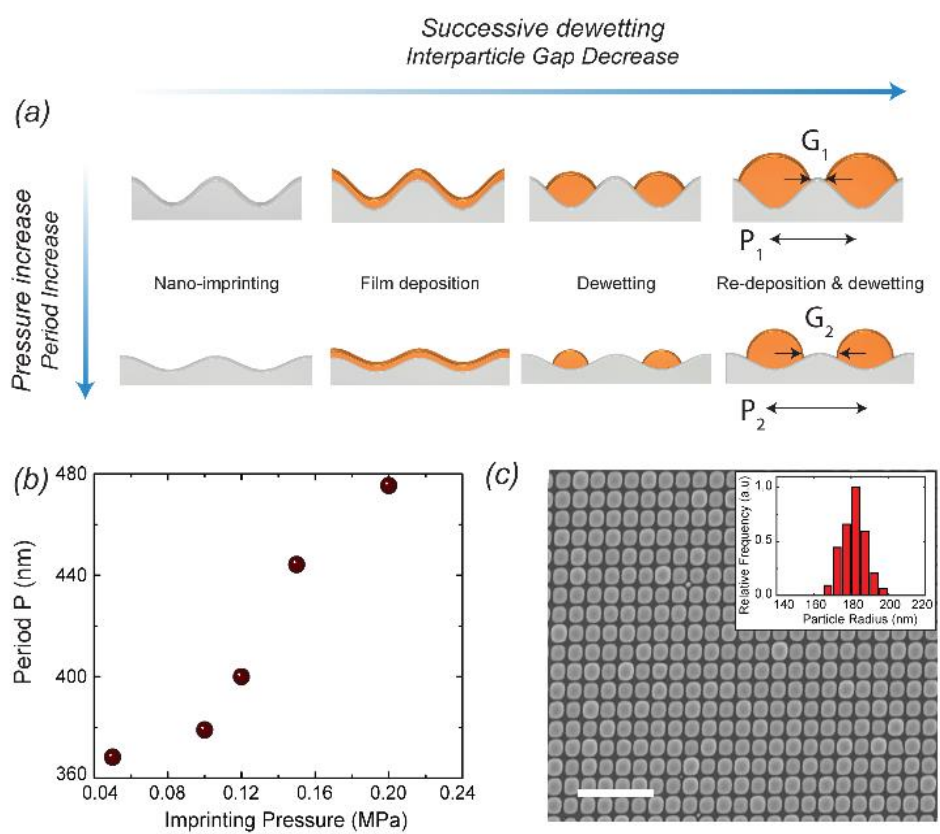

*Figure 2: Fabrication process of the metasurfaces. (a) Schematic diagram illustrating the process of fabrication. (b) Variation of the periodicity with the applied pressure. (c) Scanning electron microscopy image of dewetted Se nano-arrays obtained by successive dewetting corresponding to a total cumulative thickness of 60nm for a fixed periodicity of 400nm. The scale bar is 2 µm. The graph in inset shows the particle size distribution. A Gaussian fit provides a average radius value r=181±13 nm.*

We start by investigating a simple, yet thus far unexplored, approach to tailor the lattice period of the underlying pattern, via engineering the applied pressure during the imprinting step. Figure 2(b) shows the change in period on a bare polycarbonate (PC) textured sample when the pressure varies from 0.05 MPa to 0.2 MPa. As the applied pressure during the nanoimprinting process is increased we observe an increase in the period. The SEM images (see Figure

SI.1) suggest a well-controlled increase of the array periodicity with the applied pressure, which will be later exploited to tune the lattice resonance.

On the other hand, our process also proposes a way to control Se particle size or radius (or equivalently their spacing distance, labeled G in this work). By successively depositing and dewetting thin layers of Se of controlled thickness, it is possible to control the inter-particle distance down to approximately 10 nm (Figure S.I.2) while increasing the particle size. Controlling evaporated thickness hence provides a lever over the interparticle gap. As apparent in Figure 2(c), there is an inherent size distribution for dewetted particles. This size distribution is intricately linked to a number of factors including substrate inter-pyramid spacing, local substrate curvature and film thickness. Interestingly, the particle size distribution can be narrowed down by reducing the interparticle gap (equivalent to increasing the particle size), independently of the factors cited above. This point has already been extensively investigated in reference 39, and we refer the reader to this work for more details on this point. The particles shown in Figure 2(c) have a radial distribution of r = 181 ± 13 nm, which suggests a high degree of order. Reduced ordering proves detrimental to the resonance sharpness, characterized by the Full Width Half Maximum (FWHM). The sharpness of the resonances experimentally obtained in this work further support that the high degree of order is maintained for the considered range of particle sizes and periods.

Note that the Se particles must be tailored to support resonances at wavelengths longer than the spectral absorbing region of amorphous Se (non-negligible losses below 600 nm, see Figure S.I.3) and within the spectral range of our Ti: sapphire femtosecond laser ($\lambda$ = 650 nm – 1200 nm). The computed single particle scattering spectra (Figure S.I.3) show broad resonances, indicating that a particle size of radius above 300 nm is a prerequisite to have a resonance inducing a strong field enhancement at approximately 700 nm.

*Tunable optical properties*

   a) <u>Linear optical properties</u>

Figure 3(a) shows the change in the experimental transmission spectrum when we fix the lattice period P and vary the particle size S= 2.r = P - G by the process of successive dewetting, where r is defined as the radius of the particle. The total deposited thicknesses represented in the graph (40 nm, 50 nm and 60 nm) dictates the size of the particles and therefore also their spacing. The initial broad dip for 40 nm thickness decouples into two dips, which red shift as the thickness (and hence the size of the particles) increases, indicative of a change in the coupling regime as the particles resonance shift, as illustrated in Fig. 1. Before investigating and exploiting this effect, we experimentally study the effect of changing the lattice period P, while keeping the same deposition conditions (60 nm film thickness). As shown in Figure 3(b), it reveals the appearance of a strong asymmetric resonance, which increases when P is varied from 360 nm to 400 nm. A further increase of the period to 450 nm leads to spectral reshaping from one asymmetric Fano-type resonance to two symmetric resonances, indicative of a change in the coupling strength in the system[23]. From these experimental results, we identify a sharp resonance of FWHM of approximately 5 nm in air for a lattice period of 400 nm and 60 nm total thickness.

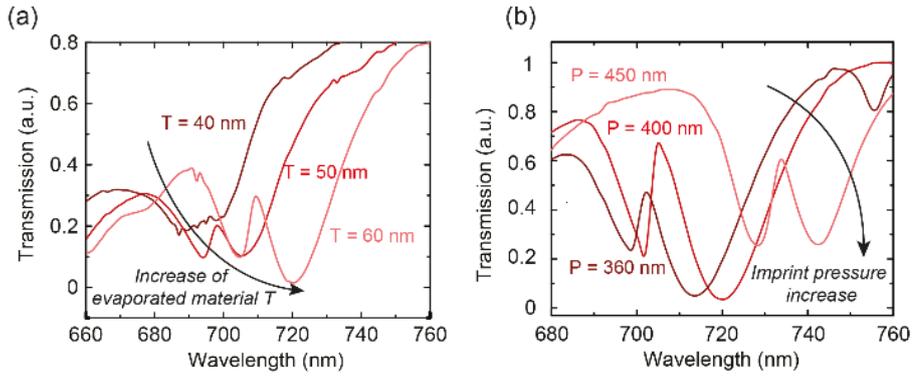

*Figure 3: Experimental transmission spectra illustrating the gap and the lattice effect. (a) Transmission spectra obtained for successive dewetting steps corresponding to an increased equivalent thickness T, maintaining the pressure during the nanoimprint at 0.1 MPa. The transmission spectrum is observed to split from a single broad resonance (at 40 nm) to two resonances (50 and 60 nm). (b) Transmission spectra for a 60 nm deposited film obtained as the pressure during the imprinting process is varied from 0.1 MPa to 0.2 MPa. As the nanoimprinted pressure is increased, the asymmetric resonance becomes sharper (0.15MPa) corresponding to increase in lattice periodicity to 400 nm. Further increases of the pressure leads to the disappearance of Fano resonance and gives rise to two broad symmetric dips (period 450 nm).*

To unveil the origin of the different resonances and their coupling, FDTD simulations were performed using the commercially available Lumerical software. The effect on the transmission spectrum of decreasing the inter-particle gap upon successive dewetting is first illustrated in Figure 4(a), keeping a fixed period P = 400nm. The inter-particle gap distance was varied from 50 to 10 nm while keeping the lattice period at 400 nm. The simulated transmission spectra (Figure 4(a)) show a sharp asymmetric dip that red-shifts on decreasing the inter-particle gap, which corresponds to experimental observations. Figure 4(b) and (c) relate the respective evolution of the electric field magnitude in the interparticle gap and the magnetic field amplitude at the particle center with wavelength. The interparticle electric field maximum (associated to the lattice resonance) both decreases and red-shifts with a reduced interparticle gap (Figure 4(b) and (d)). The evolution in Figure 4(b) corroborates the trend observed for maximal electric field intensity at resonance in the metasurface's symmetry plane (Figure 4(d)), which strongly decreases as the gap is reduced from 50 nm to 10 nm. This effect contrasts with the appearance of hot-spots at lower gap in plasmonic metasurfaces.[41,42] As shown in Figure 4(c), the magnetic field monitor (at particle center) indicates a broad magnetic resonance associated to the particle, which overlaps with another sharper magnetic peak, located at the exact position of the electric lattice resonance. This clearly suggests that Mie (magnetic) and Lattice (electric) modes interfere for the considered range of interparticle gaps.

We now turn towards the influence of the array periodicity on the optical spectrum, keeping a fixed particle size S = P - G = 350 nm. Structures with varying lattice period (from 370 nm to 450 nm, see Figures 4(e)-(g)) are illuminated with a polarized plane wave at normal incidence. The sharp asymmetric resonance apparent in the transmission spectrum (Figure 4(e)) for an initial small period increases in strength and then transforms to a symmetric dip for the highest periodicities considered in this work (450 nm). As the period increases to 600 nm (Figure S.I.4), the two resonances merge, leading to a single resonance, which disappears upon increasing the period to 800 nm. The asymmetric resonance observed in the experimental (Figure 3(a)) and simulated (Figure 4(e)) transmission spectra for

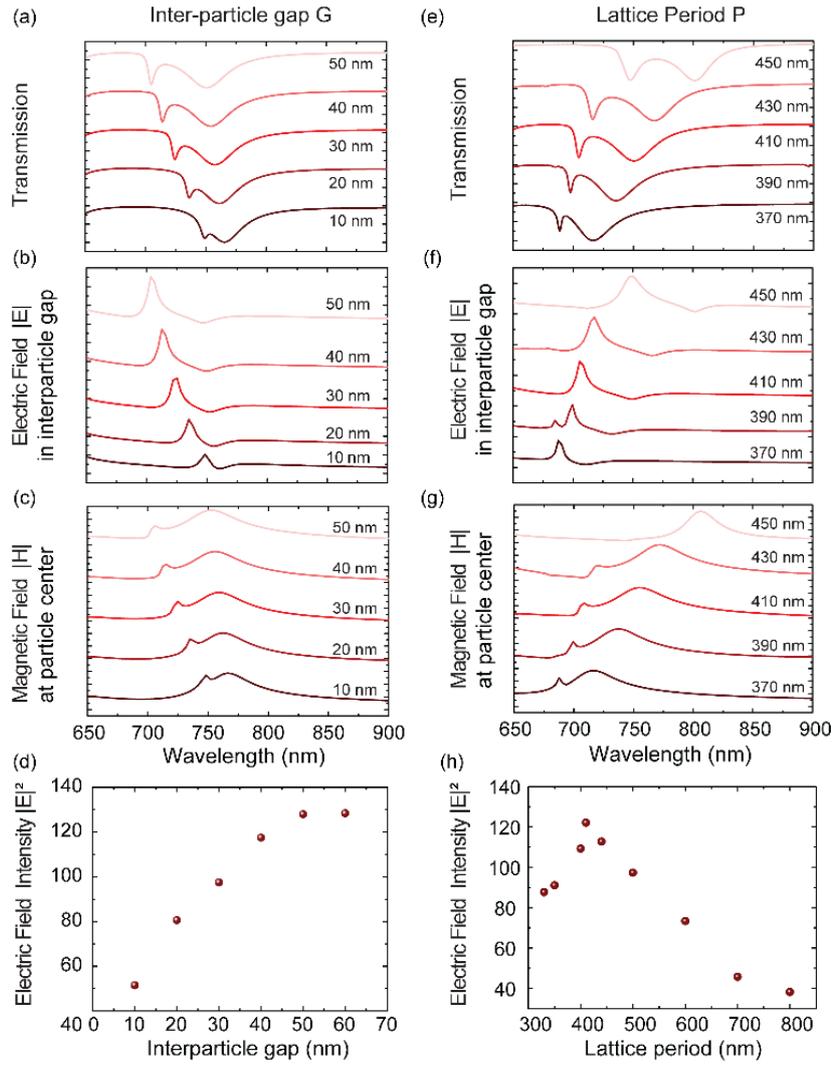

*Figure 4: Effect of the periodicity and gap on the mode properties. (a)-(d): Effect of gap / particle size for a period of 400 nm. (a) Simulated transmission spectra, (b) simulated electric field magnitude in the interparticle region, and (c) simulated magnetic field magnitude at particle center as a function of wavelength with a reduction of interparticle gap from 50 nm to 10 nm. (d) Simulated maximal electric field intensity along the unit period's symmetry plane. (e)-(h): Effect of periodicity for a fixed particle diameter of 350 nm. (e) Simulated transmission spectra, (f) simulated electric field magnitude in the interparticle region, and (g) simulated magnetic field magnitude at particle center as a function of wavelength with an increase in period from 370 nm to 400 nm. (h) Simulated maximal electric field intensity along the unit period's symmetry plane.*

a small period, is mainly attributed to the electric lattice resonance. Figure 4(f) and 4(g) show the electric and magnetic field amplitude at the interparticle gap and particle center respectively as the array period is varied from 370 nm to 450 nm. The interparticle electric field maximum (associated to the lattice resonance, Figure 4(f)) first increases with an increase in period from 370 nm to 410 nm, and decreases when the period is further expanded beyond 410 nm. This is in line with the trend observed for maximal electric field intensity at resonance, based on the electric intensity profile distribution in the metasurface's symmetry plane (Figure 4(h)), which reaches a maximal value at around 400 nm. The electric resonance associated to the lattice interferes with the broad electric resonance associated to the particle up to

430 nm period, as suggested in Figure 4(g). Increasing the period beyond 430 nm leads to a sharp drop in the field enhancement, indicating that electric lattice and electric particle modes no longer interfere.[18,43]

Thus, to obtain a high quality factor, engineering mode coupling through the fine tuning of lattice period and particle size allowed by our process is essential. For this specific substrate, texture, and dielectric material, we find that the optimal resonance overlap occurs at 700 nm for a lattice period of 400 nm and a Se particle size of 350 nm. This architecture leads to sharp resonant modes with a strong field enhancement and a high experimental Q.F. of 175 at 700 nm wavelength.

b) *Second harmonic generation*

To demonstrate the potential of our metasurfaces design capabilities, we used the previously demonstrated nano-photonic structures with optimized light-matter interactions for resonantly-enhanced SHG. Simulated linear transmission spectrum (Figure 5(a), wine-red curve) for a nanoparticle size of 350 nm and a period of 400 nm reveals a dip associated to the lattice resonance excitation. The experimental linear measurement is in line with simulation predictions, showing a corresponding dip in transmission at 700 nm (Figure 5(b), wine-red curve). This resonance leads to an enhancement of the electric field intensity by two orders of magnitude (~90x) in comparison with off-resonance spectral regions (Figure 5(b), light red markers).

SHG measurements were performed in reflection using a previously reported setup[44]. From each of the obtained scanned image, the SH intensity at particular wavelength was calculated from the mean value of those images as shown in Figure 5(c) (light red dots) when the pump wavelength is tuned between 680 nm and 780 nm. The conversion efficiency η is defined as:

$$\eta = n_{photon}^{SH} \big/ n_{photon}^{Fundamental} \quad \text{(ii)}$$

where $n_{photon}^{SH}$ and $n_{photon}^{Fundamental}$ are the photon counts of the generated second harmonic (SH) and the fundamental incident signal respectively. The intensity I is related to the photon count according to the following equation:

$$I(\lambda) = nh\frac{c}{\lambda} \quad \text{(iii)}$$

where I is the energy and λ the corresponding wavelength. We proceed to measure the re-emitted SH intensity $I_{SH}$ over a fixed integration time (10 second), as well as the intensity of the fundamental $I_{Fundamental}$ reflected from a gold mirror surface over a reduced integration time to avoid saturation (1 second). All measurements are done under normal incidence. Normalizing these powers with time and combining relations (ii) and (iii) finally provide the conversion efficiency, which is evaluated here at $10^{-6}$ at resonance and at $10^{-8}$ off-resonance. The normalized SHG power shows a peak at the asymmetric electric resonance (700 nm) due to the electromagnetic field enhancement (Figure 5(a)). When exciting at the resonant wavelength of 700 nm (i.e. maximal field enhancement), SH signal normalized with the incident power exhibits a 100 times enhancement with respect to the non-resonant region (800 nm), and almost 4 orders of magnitude enhancement as compared to the non-patterned Se thin film of comparable thickness in the off-resonant region (800 nm) (Figure S.I.5). Figure 5(d)-(c) reproduce the study done for Figure 5(a)-(c) for an increased period of 470nm with a fixed particle size of 350 nm. Based on the transmission spectrum, this situation is akin to strong coupling, and demonstrates comparatively lower field enhancement and this SHG emission as compared with the critically coupled case of Figures 5(a)-(c).

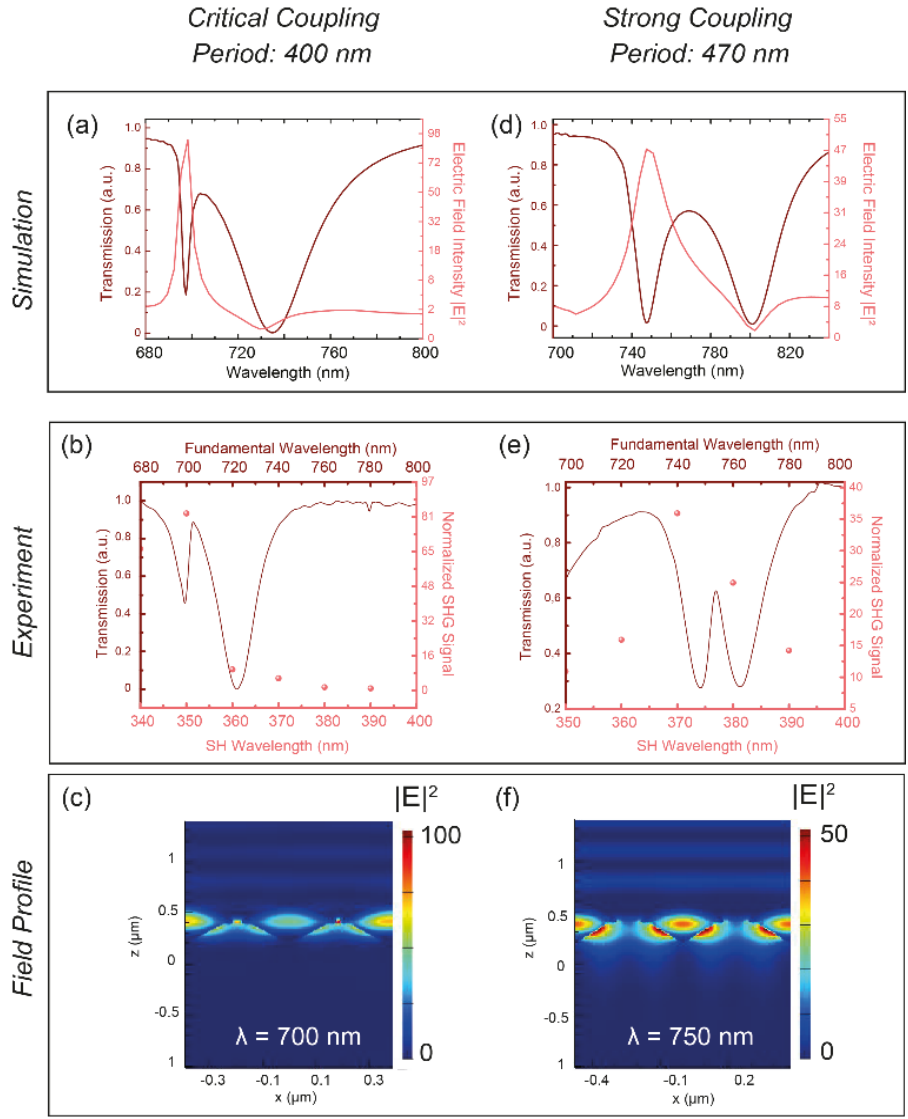

*Figure 5: Resonantly Enhanced Second Harmonic Generation in critically and strongly coupled conditions. (a)-(c): Fixed period of 400 nm and particle size of 350 nm, providing critical coupling conditions. (a) Simulated linear transmission spectrum (wine-red) and electric field intensity evaluated at the Se nanoparticle center (light red). (b) Experimental transmission spectrum (wine-red curve) and normalized second harmonic signal (with respect to the incident power, in light red dots) showing a ~90x signal enhancement. (c) Distribution of the electric field intensity in a plane perpendicular to the metasurface for an incident wavelength λ = 700 nm. (d)-(f): Fixed period of 470 nm and particle size of 350 nm, yielding a strongly coupled regime. (d) Simulated linear transmission spectrum (wine-red) and electric field intensity evaluated at the Se nanoparticle center (light red). (e) Experimental transmission spectrum (wine-red curve) and normalized second harmonic signal (with respect to the incident power, in light red dots) showing a ~ 38x signal enhancement. (f) Distribution of the electric field intensity in a plane perpendicular to the metasurface for an incident wavelength λ = 750 nm.*

The dependence of the field intensity enhancement ($I_\omega = |E|^2$) on the lattice period ($P$) for these structures, follows a quadratic law defined by:

$$I_\omega = A_o P^{-2} \quad (iv)$$

where $A_o$ is determined by the nature of meta-atoms and metasurface design (see S.I. Figure S.I.7(top)). It is worth noting that a similar quadratic dependency on the critical coupling has been observed for asymmetric structures.[45] Furthermore, the FWHM is also observed to scale quadratically with the lattice period based on numerical simulations (S.I. Figure S.I.7 (bottom)):

$$FWHM = B_o |P - P_c|^{-2} \quad (v)$$

where $B_o$ is a constant governed by the metasurface structure and $P_c$ is the critical period, which leads to critical coupling of the radiative dipole and the in-plane diffractive modes. To illustrate the SHG efficiency as a function of the period, we plotted the maximum SH signal for different periodicities (Figure S.I.8). It is observed that below and above the critical period ($P_c$), the SH intensity scales as the 4$^{th}$ power of the period ($P$). This can be rationalized from our previous numerical analogy where the fundamental intensity ($I_\omega$) scales quadratically ($P^{-2}$) above $P_c$. In a first approximation[46], the observed SH intensity is expected to scale quadratically with the fundamental field enhancement, thus demonstrating a quadratic dependency on period ($I_{2\omega} \alpha I_\omega^2 \rightarrow I_{2\omega} \alpha P^{-4}$). From our simulation, it appears that the SH signal should be at-least 8100 times. The discrepancy seen between our numerical simulation and experimental data (Figure 4) can have different origins. The total scattered SH far field intensity is proportional to the perpendicular component of the electric field intensity ($E_z(r,\omega)$). To further analyses the perpendicular ($E_z(r,\omega)$) and parallel component ($E_y(r,\omega)$) of the electric field, we plotted the mode profiles as shown in Figure S.I.9. From such mode profile plots, it suggests that the field intensity is significantly enhanced along the substrate plane, while along the z plane it is only enhanced 10 times, indicating that the normal component of the field is mainly contributing to the SH enhancement[16]. Furthermore, the detection of the SH signal in reflection mode might also be another cause of the discrepancy, since a non-negligible part of the SH wave can also propagate in the forward direction. Nevertheless, the experimental results clearly show the ability of our fabrication technique to produce efficient metasurfaces that yield both high SHG and tunable Q.F. with a low FWHM.

## Conclusions

In conclusion, with a simple process of nanoimprinting and dewetting, we developed a technique whereby optical properties of all-dielectric metasurfaces can be finely tuned whilst keeping the same original Si master. The tunability of the optical properties is possible by applying different pressures, leading to a controlled change in period. By a process of successive dewetting, increasing particle sizes were obtained, which in turn tuned the inter-particle gap. With this ability to control these two independent parameters, different optical modes with large Q.F. were observed in a simple way without undergoing any hard lithographic steps. In particular, we have shown that by this programmable process, field enhancement could be increased up to a 100 times. Such enhancement of field could give rise to SHG with a tailored FWHM using an amorphous structure. We have experimentally and numerically verified that the performance of SH intensity can be significantly improved by proper tuning the different optical modes in our metasurface. We demonstrated a SH conversion efficiency that is significantly larger compared to plasmonic metasurfaces and Silicon. Our approach could also be generally applied to III-V semiconductor in the UV or VUV regions, where sharp SH signals are required for bio-molecule sensing application. Such general design approach could indeed pave the way for fabrication of nonlinear meta-devices, fast optical switching, frequency conversion, and light-based communication.

**Conflicts of interest**

The authors declare no conflicts of interest.

**Acknowledgements**

The authors acknowledge the European Research Council for funding support (ERC starting grant 679211 'FLOWTONICS' and ERC-2015-AdG-695206 Nanofactory) and the Swiss National Science Foundation (project 200020_153662). The authors would like to acknowledge Dr. Christian Santschi and Thierry Laroche for their help in the non-linear measurements.

# Supplementary Informations

*1. Sample Fabrication process*

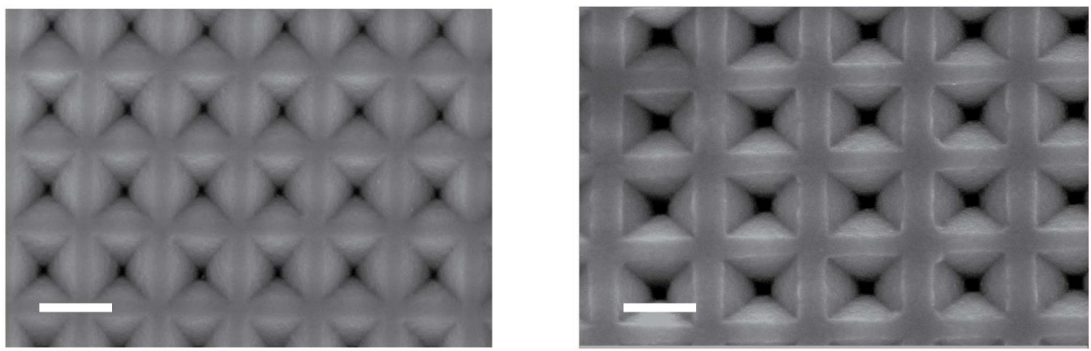

Figure S.I.1. SEM image illustrating the effect of applied pressure during nanoimprinting (left) period 360 nm (applied Pressure 0.05 MPa) (right) period 480 nm (applied pressure 0.2MPa) . Scale Bar 350 nm.

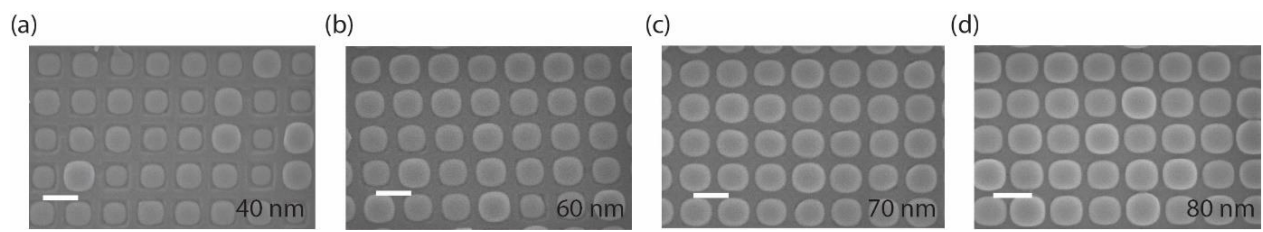

Figure S.I.2. Successive dewetting process. SEM images (a) to (d) illustrating that by the process of successive dewetting (deposition and dewetting of successive thin films) the inter-particle gap and particle size can be tuned. Inset shows the total deposited thickness of the film.

*2. Optical constant of Selenium*

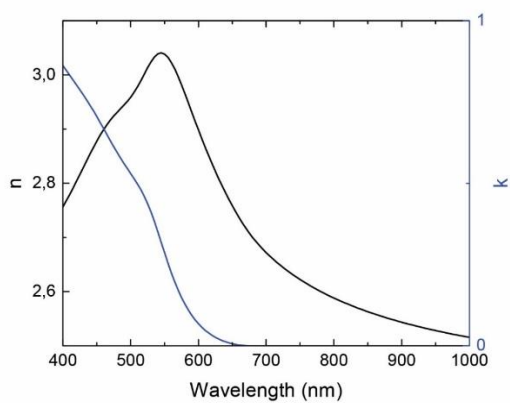

Figure S.I.3. Optical constant of Selenium obtained by ellipsometry

## 3. Single particle scattering analysis

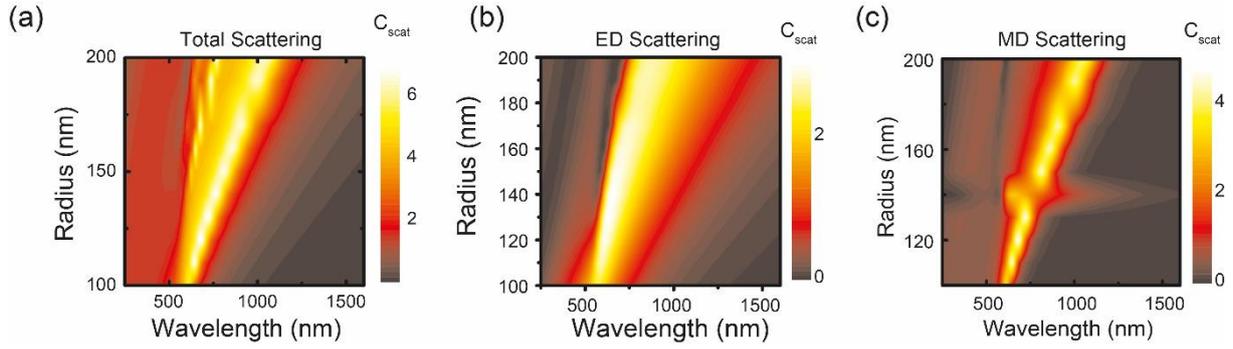

*Figure S.I.4. Single particle numerical analysis as a function of radius (a) multipole expansion of scattering spectra (b) electric dipole (c) magnetic dipole evaluated with Mie theory for Selenium (Se).*

Mie theory was used to calculate the single particle scattering cross-section. Scattering cross-section of a single homogeneous sphere is given by:

$$C_{scat} = \frac{2\pi}{k^2}\sum_n (|a_n|^2 + |b_n|^2)$$

Where $k$ is the wavenumber, $a_n$, $b_n$ are the nth order of electric and magnetic Mie coefficients respectively; and n=1 and 2 indicates the dipole and quadruple moments respectively. Figure 1(a) shows the calculated $C_{scat}$ as a function of particle radius and wavelength, revealing a red-shift for both the electric dipole (ED) and the magnetic dipole (MD) resonance. In a sharp contrast to the broad ED, MD exhibits a narrower linewidth with a larger contribution to the total extinction cross-section than the ED resonance. The main contribution of the scattering spectra thus comes from the MD for larger particle size. Importantly, both the ED and MD resonances lie above the lossless region of the bulk selenium for a particle size of 300 nm, satisfying design criteria.

## 4. Effect of mode coupling by lattice tuning

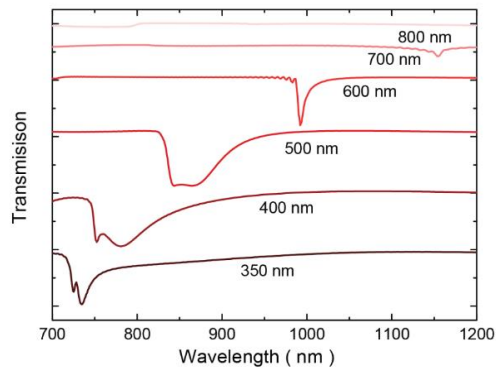

*Figure S.I.5. Simulation of transmission illustrating the dip positions with increasing periodicities.*

## 5. Field profile distribution at critical coupling

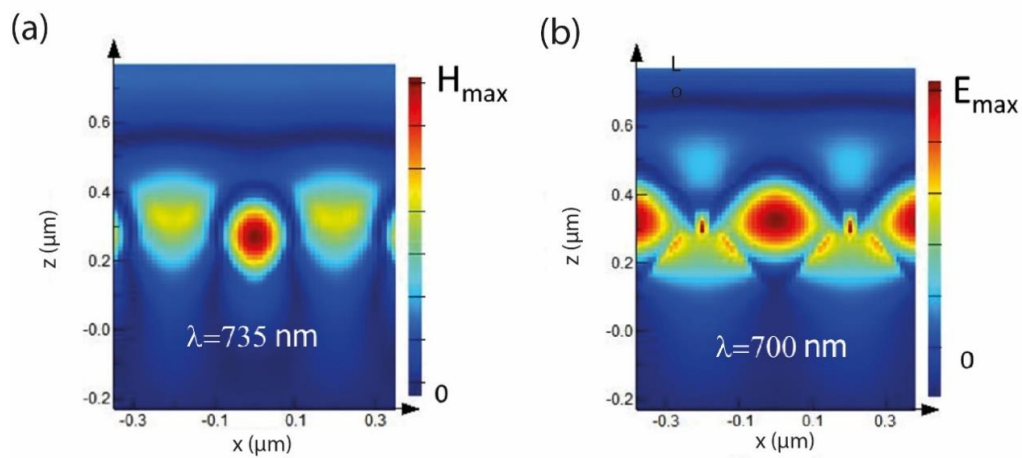

*Figure S.I.6. (a) Magnetic and (b) electric plot at their corresponding resonance for a 400 nm period.*

## 6. Power law dependency of SHG

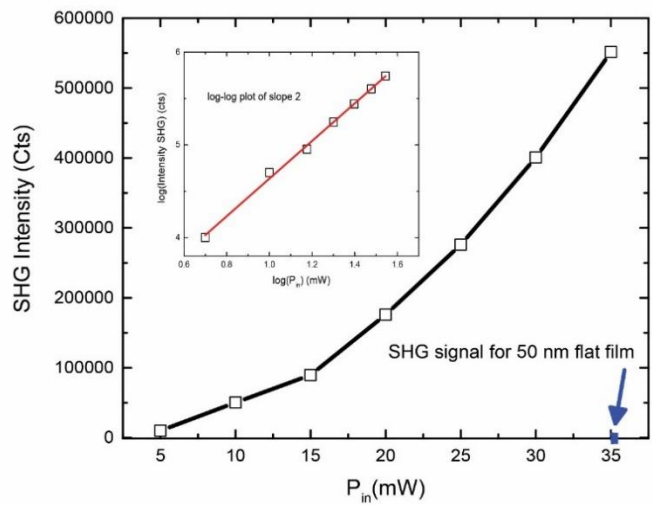

*Figure S.I.7. Power law dependency of the emitted signal at 400 nm when the fundamental wavelength is 800 nm. Inset shows the log-log plot of the curve showing a linear fit of slope 2. SHG intensity of the metasurface due to fundamental field enhancement is at least 4 orders of magnitude higher than the thin film.*

## 7. SHG Spectrum at Resonant and non resonant region

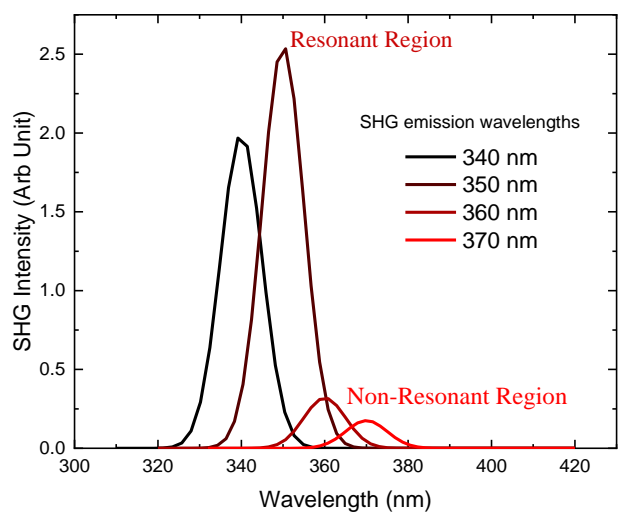

*Figure S.I.8. SHG emitted spectrum at resonant and non-resonant region*

## 8. Polarization dependency of emitted SHG

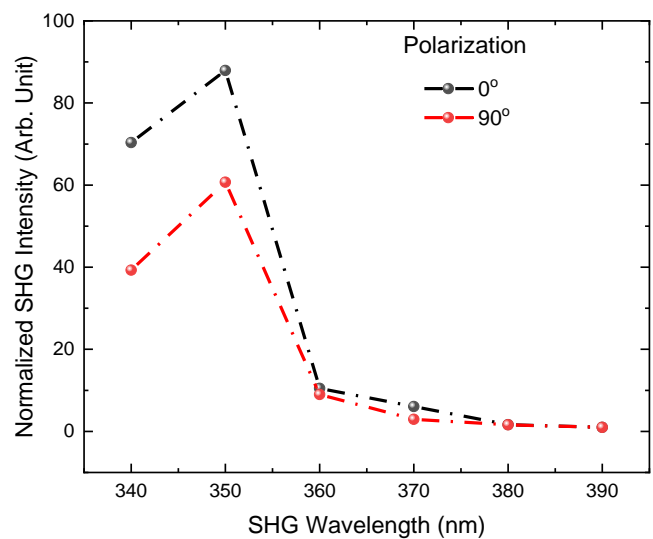

*Figure S.I.9. Polarization dependency of SHG emission*

## 9. Quadratic dependency of the Field enhancement

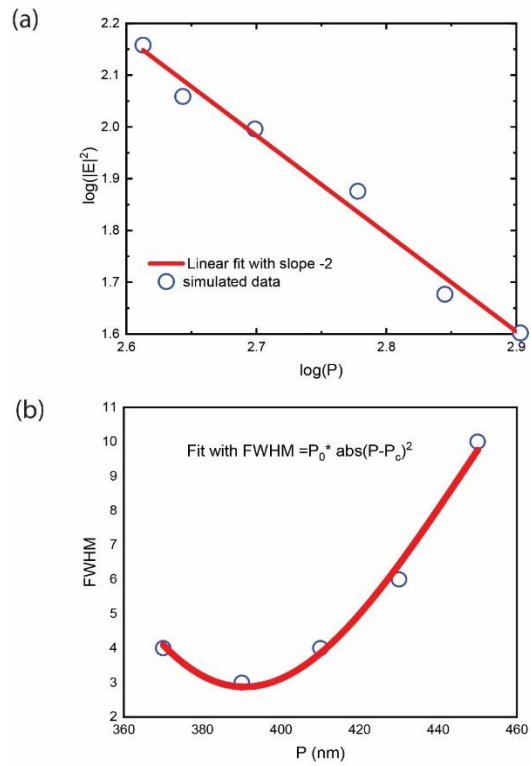

Figure S.I.10. (a) Quadratic dependency of electric field Intensity (E) on lattice period (b) Plot of Full Width Half-Maximum (FWHM) with lattice period showing the Minimum FWHM (corresponding to maximum Q.F.) with lattice period (P).

## 8. SH-signal 4th Power dependency with lattice period

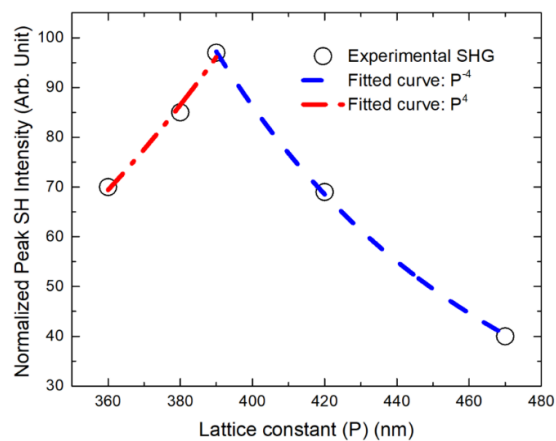

Figure S.I.11. Plot illustrating the $4^{th}$ power dependency of SH intensity as a function of lattice period (P). The SH intensity is seen to reach maximum at the critical lattice period attributed to critical coupling.

## 9. Electric field component distribution

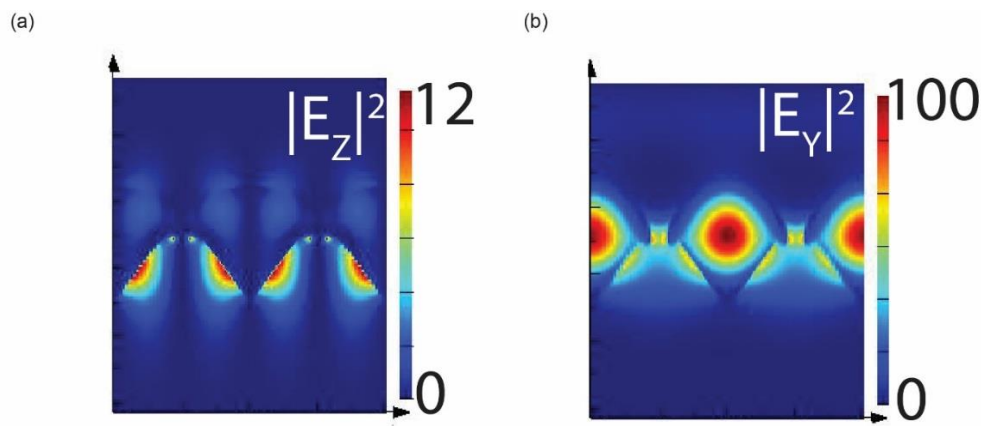

*Figure S.I.12. Electric Field component distribution (a) in z-direction showing near 10 times field intensity enhancement in the z-direction. (b) 100 times enhancement in the y direction.*

## References

1  A. Arbabi, Y. Horie, M. Bagheri and A. Faraon, *Nat. Nanotechnol.*, 2015, **10**, 937–943.

2  D. G. Baranov, D. A. Zuev, S. I. Lepeshov, O. V. Kotov, A. E. Krasnok, A. B. Evlyukhin and B. N. Chichkov, *Optica*, 2017, **4**, 814.

3  M. Decker and I. Staude, *J. Opt.*, 2016, **18**, 103001.

4  M. Decker, I. Staude, M. Falkner, J. Dominguez, D. N. Neshev, I. Brener, T. Pertsch and Y. S. Kivshar, *Adv. Opt. Mater.*, 2015, **3**, 813–820.

5  N. Yu and F. Capasso, *Nat. Mater.*, 2014, **13**, 139–150.

6  B. Metzger, M. Hentschel and H. Giessen, *Nano Lett.*, 2017, **17**, 1931–1937.

7  D. Smirnova, A. I. Smirnov and Y. S. Kivshar, *Phys. Rev. A*, 2018, **97**, 013807.

8  P. P. Vabishchevich, S. Liu, M. B. Sinclair, G. A. Keeler, G. M. Peake and I. Brener, *ACS Photonics*, 2018, **5**, 1685–1690.

9  J. D. Sautter, L. Xu, A. E. Miroshnichenko, M. Lysevych, I. Volkovskaya, D. A. Smirnova, R. Camacho-Morales, K. Zangeneh Kamali, F. Karouta, K. Vora, H. H. Tan, M. Kauranen, I. Staude, C. Jagadish, D. N. Neshev and M. Rahmani, *Nano Lett.*, 2019, **19**, 3905–3911.

10  S. V. Makarov, M. I. Petrov, U. Zywietz, V. Milichko, D. Zuev, N. Lopanitsyna, A. Kuksin, I. Mukhin, G. Zograf, E. Ubyivovk, D. A. Smirnova, S. Starikov, B. N. Chichkov and Y. S. Kivshar, *Nano Lett.*, 2017, **17**, 3047–3053.

11  J. Butet, P.-F. Brevet and O. J. F. Martin, *ACS Nano*, 2015, **9**, 10545–10562.

12  M. J. Huttunen, R. Czaplicki and M. Kauranen, *J. Nonlinear Opt. Phys. Mater.*, 2019, **28**, 1950001.

13  M. Hentschel, B. Metzger, B. Knabe, K. Buse and H. Giessen, *Beilstein J. Nanotechnol.*, 2016, **7**, 111–120.

14  M. Parry, A. Komar, B. Hopkins, S. Campione, S. Liu, A. E. Miroshnichenko, J. Nogan, M. B. Sinclair, I. Brener and D. N. Neshev, *Appl. Phys. Lett.*, 2017, **111**, 053102.

15  M. V. Rybin, K. L. Koshelev, Z. F. Sadrieva, K. B. Samusev, A. A. Bogdanov, M. F. Limonov and Y. S. Kivshar, *Phys. Rev. Lett.*, 2017, **119**, 243901.

16  F. Wang, A. B. F. Martinson and H. Harutyunyan, *ACS Photonics*, 2017, **4**, 1188–1194.

17  J. T. Collins, D. C. Hooper, A. G. Mark, C. Kuppe and V. K. Valev, *ACS Nano*, 2018, **12**, 5445–5451.

18  D. C. Hooper, C. Kuppe, D. Wang, W. Wang, J. Guan, T. W. Odom and V. K. Valev, *Nano Lett.*, 2019, **19**, 165–172.

19  M. H. Lee, M. D. Huntington, W. Zhou, J.-C. Yang and T. W. Odom, *Nano Lett.*, 2011, **11**, 311–315.

20  V. G. Kravets, A. V. Kabashin, W. L. Barnes and A. N. Grigorenko, *Chem. Rev.*, 2018, **118**, 5912–5951.

21  A. Hessel and A. A. Oliner, *Appl. Opt.*, 1965, **4**, 1275.

22  A. Maradudin, I. Simonsen, J. Polanco, and R. Fitzgerald, *Jour. Opt.*, 2016, **18**, 024004.

23  B. Gallinet, T. Siegfried, H. Sigg, P. Nordlander and O. J. F. Martin, *Nano Lett.*, 2013, **13**, 497–503.

24  Y. Chu and K. B. Crozier, *Opt. Lett.*, 2009, **34**, 244.

25  A. Christ, Y. Ekinci, H. H. Solak, N. A. Gippius, S. G. Tikhodeev and O. J. F. Martin, *Phys. Rev. B - Condens. Matter Mater. Phys.*, 2007, **76**, 201405.

26  P. Törmä and W. L. Barnes, *Reports Prog. Phys.*, 2015, **78**, 013901.

27  L. Michaeli, S. Keren-Zur, O. Avayu, H. Suchowski and T. Ellenbogen, *Phys. Rev. Lett.*, 2017, **118**, 243904.

28  R. Czaplicki, A. Kiviniemi, M. J. Huttunen, X. Zang, T. Stolt, I. Vartiainen, J. Butet, M. Kuittinen, O. J. F. Martin and M. Kauranen, *Nano Lett.*, 2018, **18**, 7709–7714.

29  A. Kiselev, G. D. Bernasconi and O. J. F. Martin, *Opt. Express*, 2019, **27**, 38708.

30  L. Liu, J. Zhang, M. A. Badshah, L. Dong, J. Li, S. Kim and M. Lu, *Sci. Rep.*, 2016, **6**, 22445.

31  A. Yang, A. J. Hryn, M. R. Bourgeois, W.-K. Lee, J. Hu, G. C. Schatz and T. W. Odom, *Proc. Natl. Acad. Sci. U. S. A.*, 2016, **113**, 14201–14206.

32  B. J. Eggleton, B. Luther-Davies and K. Richardson, *Nat. Photonics*, 2011, **5**, 141–148.

33  H. Lin, Y. Song, Y. Huang, D. Kita, S. Deckoff-Jones, K. Wang, L. Li, J. Li, H. Zheng, Z. Luo, H. Wang, S. Novak, A. Yadav, C.-C. Huang, R.-J. Shiue, D. Englund, T. Gu, D. Hewak, K. Richardson, J. Kong and J. Hu, *Nat. Photonics*, 2017, **11**, 798–805.

34  W. Yan, T. Nguyen-Dang, C. Cayron, T. Das Gupta, A. G. Page, Y. Qu and F. Sorin, *Opt. Mater. Express*, 2017, **7**, 1388.

35  W. Yan, Y. Qu, T. Das Gupta, A. Darga, D. T. Nguyên, A. G. Page, M. Rossi, M. Ceriotti and F. Sorin, *Adv. Mater.*, 2017, **29**, 1700681.

36  Y. Zou, L. Moreel, H. Lin, J. Zhou, L. Li, S. Danto, J. D. Musgraves, E. Koontz, K. Richardson, K. D. Dobson, R. Birkmire and J. Hu, *Adv. Opt. Mater.*, 2014, **2**, 759–764.

37  M. Guignard, V. Nazabal, J. Troles, F. Smektala, H. Zeghlache, Y. Quiquempois, A. Kudlinski and G. Martinelli, *Opt. Express*, 2005, **13**, 789.

38  C. R. Ma, J. H. Yan, Y. M. Wei and G. W. Yang, *Nanotechnology*, 2016, **27**, 425206.

39  T. Das Gupta, L. Martin-Monier, W. Yan, A. Le Bris, T. Nguyen-Dang, A. G. Page, K.-T. Ho, F. Yesilköy, H. Altug, Y. Qu and F. Sorin, *Nat. Nanotechnol.*, 2019, **14**, 320–327.

40  N. Sultanova, S. Kasarova and I. Nikolov, in *Acta Physica Polonica A*, Polish Academy of Sciences, 2009, vol. 116, pp. 585–587.

41  C. Readman, B. De Nijs, I. Szabó, A. Demetriadou, R. Greenhalgh, C. Durkan, E. Rosta, O. A. Scherman and J. J. Baumberg, *Nano Lett.*, 2019, **19**, 2051–2058.

42  A. Manjavacas, L. Zundel and S. Sanders, *ACS Nano*, 2019, **13**, 10682–10693.

43  M. J. Huttunen, P. Rasekh, R. W. Boyd and K. Dolgaleva, *Phys. Rev. A*, 2018, **97**, 053817.

44  K. Y. Yang, J. Butet, C. Yan, G. D. Bernasconi and O. J. F. Martin, *ACS Photonics*, 2017, **4**, 1522–1530.

45  K. Koshelev, Y. Tang, K. Li, D.-Y. Choi, G. Li and Y. Kivshar, *ACS Photonics*, 2019, **6**, 1639–1644.

46  J. Wang, J. Butet, A.-L. Baudrion, A. Horrer, G. Lévêque, O. J. F. Martin, A. J. Meixner, M. Fleischer, P.-M. Adam, A. Horneber and D. Zhang, *J. Phys. Chem. C*, 2016, **120**, 17699–17710.